%% file: main.tex
\documentclass[%
 reprint,
superscriptaddress,
nofootinbib,
 amsmath,amssymb,
 aps,
]{revtex4-2}

\usepackage[english]{babel}
\usepackage[letterpaper,top=2cm,bottom=2cm,left=2.0cm,right=2.0cm,marginparwidth=1.75cm]{geometry}

\usepackage{amsmath}
\usepackage{graphicx}
\usepackage[colorlinks=true, allcolors=blue]{hyperref}
\usepackage{dcolumn}
\usepackage{bm}
\usepackage{booktabs}
\usepackage{xcolor,colortbl}
\usepackage[normalem]{ulem}

\begin{document}

\title{Interferometry Radii at RHIC BES Energies within the integrated HydroKinetic Model}

\author{Vladislav Naboka}
\email{nvlad1@ukr.net}
 \affiliation{Bogolyubov Institute for Theoretical Physics, Metrolohichna  14b, 03143 Kyiv,  Ukraine}
 
\author{Yuri Sinyukov}%
\email{Yu.Sinyukov@gmail.com}
\affiliation{Bogolyubov Institute for Theoretical Physics, Metrolohichna  14b, 03143 Kyiv,  Ukraine}
\affiliation{%
 Warsaw University of Technology, Faculty of Physics, Koszykowa 75, 00-662 Warsaw, Poland\\
}%

\author{Musfer Adzhymambetov}
\affiliation{Bogolyubov Institute for Theoretical Physics, Metrolohichna  14b, 03143 Kyiv,  Ukraine}

\author{Hanna Zbroszczyk}
\affiliation{%
 Warsaw University of Technology, Faculty of Physics, Koszykowa 75, 00-662 Warsaw, Poland\\
}%

\date{\today}

\begin{abstract}
The work is devoted to research of pion femtoscopic correlations in relativistic heavy-ion collisions across the RHIC Beam Energy Scan range using the extended integrated HydroKinetic Model (iHKMe). The model provides a comprehensive description of the system’s dynamical evolution, starting from the initial collision state of colliding nuclei, passing through a possible thermalization process and hydrodynamic expansion to the hadron interacting cascade and formation of the observed particle spectra. The model smoothly couples all these stages of the matter evolution, ensuring a smooth transition between the stages. A primary focus of the current work, in contrast to the similar investigation within iHKM for high energies, concentrated in the energy region where extended nuclear overlap times and incomplete thermalization significantly influence the system’s expansion comparing with very high energies. We extract the three-dimensional interferometry radii ($R_{\text{out}}$, $R_{\text{side}}$, $R_{\text{long}}$) in the region from 7.7 to 39 GeV per nucleon pair and evaluate their sensitivity to the features of the equation of state (EoS), specifically comparing crossover and first-order phase transition scenarios. The model demonstrates good agreement with experimental measurements in the crossover case. As for the first-order phase transition scenario, the $R_{\text{long}}$ component, at the optimal model parameters for particle spectra, is noticeably higher than the experimental data, and the difference is more pronounced with the growth of collision energy. Such a behavior is caused by the increased system’s lifetime during the mixed-phase stage. The corresponding analysis within iHKMe below $\sqrt{s_{NN}} = 7.7$ GeV will be presented within a separate investigation.

\end{abstract}

\maketitle

\section{Introduction}

\input{sections/1_introduction}

\section{Integrated hydrokinetic model}\label{sec:model}

\input{sections/2_iHKMe}

\section{Results}\label{sec:results}

\input{sections/3_results}

\section{Summary}\label{sec:summary}

\input{sections/4_summary}

\bibliography{bibliography}

\end{document}

%% file: sections/1_introduction.tex
Relativistic heavy-ion collisions have long served as a powerful tool for exploring the properties of strongly interacting matter under extreme conditions of temperature, baryon density, and energy density. Such conditions, achieved at facilities like the Relativistic Heavy Ion Collider (RHIC) and the Large Hadron Collider (LHC), are believed to resemble those that existed in the early Universe, microseconds after the Big Bang \cite{Rafelski2013}. A central objective of these experiments is to study the formation and evolution of the quark-gluon plasma (QGP) and to investigate the nature of the transition between deconfined and confined phases of quantum chromodynamics (QCD) matter \cite{Adams2005, Adcox2005}. \\

In recent years, experimental programs have increasingly turned their attention to lower collision energies—particularly through the RHIC Beam Energy Scan (BES) program—where it has become evident that the underlying physical conditions differ significantly from those at top RHIC or LHC energies \cite{Odyniec2013}. At lower center-of-mass energies ($\sqrt{s_{NN}}$), the nuclear overlap time ($\tau_{\text{overlap}} \approx 1$–$4$ fm/$c$) is no longer negligible compared to the prethermalization timescale, net baryon densities are substantial, and full thermalization may not be achieved. Moreover, the phase transition between hadronic and partonic matter is expected to exhibit critical behavior, potentially indicating the presence of a critical endpoint in the QCD phase diagram \cite{Stephanov1998, Stephanov1999}. \\

A key method for probing the space-time structure of the particle-emitting source is two-particle interferometry \cite{Goldhaber1960, Kopylov1974}, commonly referred to now as correlation femtoscopy  (CF) or Hanbury Brown–Twiss analysis (HBT) \cite{HBT}. (The first one is correctly associated with quantum correlation effects between momenta of identical bosons emitted from extremely close space-time points during  the  particle/nuclear collision process, the second one originates from investigation of intensity correlations of classical electromagnetic fields detected simultaneously from different parts of a far star. The  principal similarity, based on properties of boson fields, and the essential  difference  between the two aforementioned original methods are analysed in detail in \cite{Sinyukov:2013Femtoscopy}) Measurements of identical particle correlations enable the extraction of interferometry radii, which are sensitive to both the geometric size and dynamic evolution of the system. Accurately modeling these observables requires a realistic description of the full system evolution, including pre-equilibrium dynamics, thermalization, hydrodynamic expansion, and freeze-out processes. \\


The energy dependence of the femtoscopic radii has by now been mapped continuously from AGS and SPS energies through top RHIC \cite{STAR:2015} and LHC energies \cite{ALICE:2011dyt}, and, more recently, extended down to the lowest energies currently reachable in the laboratory by the HADES collaboration at $\sqrt{s_{NN}} = 2.3$--$2.4$~GeV \cite{HADES:2018gop}. Interestingly, the HADES data show only a mild variation of the source radii over nearly three orders of magnitude in collision energy compared to LHC, while at the same time exhibiting a pronounced charge-sign splitting at low transverse momentum, likely driven by the Coulomb interaction between the pion pair and the dense residual charge of the fireball.

Once the analysis is extended away from mid-rapidity or to non-central collisions, the correlation function acquires additional structure: the transverse and longitudinal radii pick up a dependence on pair rapidity \cite{ATLAS:2017shk}, nonzero cross terms $R_{ol}^2$ and  $R_{sl}^2$ appear and encode the tilt of the emitting source relative to the beam axis \cite{Lisa:2000ip, Greifenhagen:2020dqh}, and the correlation function itself develops non-Gaussian features not captured by the simple three-parameter fit \cite{STAR:2015}. These effects are well documented and physically rich, but they are also, by construction, higher-order corrections to the mid-rapidity, azimuthally-integrated radii that set the basic space-time scale of the source. Since our goal here is to test the sensitivity of that basic scale -- specifically $R_{\text{out}}$, $R_{\text{side}}$, and $R_{\text{long}}$ at mid-rapidity -- to the order of the QCD phase transition within iHKMe, we work within the standard three-dimensional Gaussian framework, where the cross terms vanish by symmetry and non-Gaussian corrections are subdominant; a differential analysis along these directions is a natural next step once the leading-order picture presented here is established.

Several studies have utilized correlation to investigate the space-time structure of particle emission in heavy-ion collisions.\cite{Stefaniak:2017epjconf, Brzezinski:2016WPCF, Li:2023SciChinaPhysMechAstron, Zhang:2017, Batyuk:2018}  In the EPOS-based study \cite{Stefaniak:2017epjconf}, femtoscopic pion correlations were simulated for Au+Au collisions at BES energies, and the extracted HBT radii were compared to experimental data. THERMINATOR-based studies provide a statistical framework, replicating pion correlation patterns across varying collision energies \cite{Brzezinski:2016WPCF}. In the work by Li et al. \cite{Li:2023SciChinaPhysMechAstron}, HBT radii were researched for lower energies $\sqrt{s_{NN}}=2.4 - 7.7$~GeV. It emphasizes the sensitivity of HBT radii, particularly $R_O/R_S$ and $R_O^2-R_S^2$, to the equation of state (EoS), revealing signatures of phase transitions in the QCD matter. These models, both microscopic and macroscopic, highlight differing emission durations and source sizes as possible indicators of the QGP phase and critical point. Comparative analyses between cascade and hybrid models further stress the importance of EoS stiffness. \\

The integrated HydroKinetic Model (iHKM) \cite{Naboka:2015qra, Akkelin:2009nz, Naboka:2014eha} was developed to simulate the entire evolution of ultrarelativistic nuclear collisions and has successfully reproduced a wide range of observables at top RHIC and LHC energies. It incorporates all major stages of the collision process: initial state generation, pre-equilibrium relaxation, viscous hydrodynamic evolution, particlization, and a subsequent hadronic cascade. Previous studies have shown that iHKM provides a robust description of particle spectra, flow coefficients, and femtoscopic radii at high energies \cite{Shapoval:2020nec}. \\

To address the challenges associated with modeling collisions at lower beam energies, a recently {\it extended} version of iHKM (iHKM{\it e}) has been developed \cite{Adzhymambetov:2024zzz}. In this work, we apply the updated framework to collisions in the GeV energy range, where the system experiences longer nuclear overlap times, slower thermalization, and possibly incomplete QGP formation. Unlike the original formulation \cite{Naboka:2015qra}, the extended model incorporates the effects of high net baryon density into both the equation of state and the dynamical evolution of conserved charge currents. Additionally, it features a smooth transition from the initial non-equilibrium phase, modeled via UrQMD transport dynamics \cite{urqmd1, urqmd2}, to viscous hydrodynamics during the pre-equilibrium relaxation stage. \\

In our previous work \cite{Rathod:2025gvj}, we calibrated the model parameters for RHIC BES energies based on heuristic arguments and transverse momentum spectra analyses. Two scenarios for the equation of state of strongly interacting matter were considered: one assuming a smooth crossover transition \cite{Steinheimer:2010ib}, and the other featuring a first-order phase transition between hadronic matter and the QGP \cite{Kolb:2003dz}. Interestingly, both EoS scenarios yielded comparably good agreement with the measured light hadron spectra. At higher BES energies (14.5 GeV and above), most model parameters required minimal adjustment, as the rapid system 3D-expansion reduces sensitivity to the details of the phase transition. \\

However, transverse momentum spectra alone provide limited information about the space-time characteristics of the particle-emitting source. Consequently, more sensitive observables are needed to further constrain the dynamics and probe the nature of the QCD phase transition. In this paper, we focus on the pion Bose–Einstein correlations, which are also known as correlation femtoscopy or HBT analyses.   The extracted interferometry radii reflect the system’s spatial and temporal evolution and are influenced by collective flow and expansion dynamics. We perform three-dimensional Gaussian fits to the correlation functions to extract these radii and assess their sensitivity to the underlying physics. We present results from three-dimensional fits to the correlation functions, assuming a Gaussian source shape characterized by the lengths of homogeneity: \( R_{\text{out}} \), \( R_{\text{side}} \), and \( R_{\text{long}} \).

The structure of this paper is as follows: In Sec.~\ref{sec:model}, we provide a detailed description of the extended iHKM framework as adapted for RHIC BES energies. In Sec.~\ref{sec:results}, we discuss the calibration of model parameters and present our results for interferometry radii. Finally, in Sec.~\ref{sec:summary}, we summarize our findings and outline potential directions for future research.

%% file: sections/2_iHKMe.tex
\subsection{Model description}

The integrated HydroKinetic Model (iHKM) was developed to simulate ultrarelativistic heavy-ion collisions \cite{Naboka:2015qra}. It has successfully reproduced a broad range of observables, including hadron yields, transverse momentum spectra, flow coefficients, and interferometry radii \cite{femto2016, proton-lambda2015, photons2018, photons2020, femto2018, femto2019, femto2020, Shapoval:2020nec}. Recently, the model has been extended to lower collision energies \cite{Adzhymambetov:2024zzz}, making it applicable to RHIC Beam Energy Scan (BES) and upcoming FAIR experiments.

This section briefly outlines the five stages of the iHKM framework: initial state generation, pre-equilibrium relaxation, viscous hydrodynamic evolution, particlization, and the subsequent hadronic cascade. \\

\subsubsection{Initial state generation}

Accurate modeling of ultrarelativistic heavy-ion collisions begins with generating realistic initial conditions. At high collision energies, we use the Monte Carlo Glauber model \cite{alvioli2009}. At lower energies, however, the nuclear overlap time becomes comparable to the thermalization timescale — on the order of a few fm/c — necessitating a dynamical treatment of initial state formation.

To capture this early, non-equilibrium stage, we employ the Ultrarelativistic Quantum Molecular Dynamics (UrQMD) model \cite{urqmd1, urqmd2}, a microscopic transport approach that describes the collision in terms of hadrons and resonances, with string excitation and fragmentation governing particle production at the higher end of the BES energy range. We perform full UrQMD simulations and use them, following the procedure of \cite{Rathod:2025gvj}, to split events into centrality classes; the selected events then serve to construct the energy-momentum tensor and baryon current that provide the initial conditions for the subsequent hydrodynamic stage. \\

The UrQMD energy-momentum tensor and baryon current are constructed on a sequence of proper-time layers $\tau_j$, each separated by $\Delta\tau$, the time step used in the subsequent hydrodynamic evolution; a layer $\tau_j$ collects the momenta and coordinates of all hadrons and resonances present in a given UrQMD event within $\tau_j\pm\Delta\tau/2$. Since UrQMD propagates particles in Cartesian coordinates $(t,x,y,z)$ rather than in $(\tau,x,y,\eta_s)$, a particle is assigned to layer $\tau_j$ whenever its trajectory satisfies $|\sqrt{t^2-z^2}-\tau_j|\le\Delta\tau/2$. The particles collected in this way, event by event, are then used to build the tensor and current, with each particle treated as a localized source smeared with a Gaussian profile,
\begin{equation}\label{Turqmd}
T^{\mu\nu}_{\text{urqmd}}(x) = \sum_{i} \frac{p_i^\mu p_i^\nu}{p_i^0}\, G_R(x-x_i) ,
\end{equation}
\begin{equation}\label{Jurqmd}
J^{\mu}_{\text{urqmd}}(x) = \sum_{i} B_i\,\frac{p_i^\mu}{p_i^0}\, G_R(x-x_i) ,
\end{equation}
where the sum runs over all particles $i$ collected on the layer $\tau_j$ in a given UrQMD event, $B_i$ is the baryon charge of particle $i$, and $G_R$ is a normalized Gaussian smearing kernel of width $R$ acting in the plane transverse to the beam axis and along the space-time rapidity direction \cite{Adzhymambetov:2024zzz}.

\subsubsection{Relaxation stage}

The matter produced in the early UrQMD evolution is far from equilibrium and requires thermalization before hydrodynamics can be applied. This motivates the introduction of a relaxation stage, during which the system evolves gradually toward local equilibrium. The approach is inspired by the relaxation-time approximation to the Boltzmann equation \cite{Akkelin:2009nz, Naboka:2014eha}. The relaxation stage starts at a proper time $\tau_0$ and lasts until the time $\tau_{\text{th}}$, both being free parameters of the model.

The total energy-momentum tensor and baryon charge current are taken in the form
\begin{equation}\label{eq:tensorSplit1}
    T^{\mu\nu}_{\text{total}}(x) = {\cal P}(\tau)\,T^{\mu\nu}_{\text{urqmd}} + \left(1-{\cal P}(\tau)\right)\,T^{\mu\nu}_{\text{hydro}}\,,
\end{equation}
thus representing a sum of two components: a nonequilibrium component constructed from UrQMD, Eq.~(\ref{Turqmd}), and an equilibrium one, given by the hydrodynamic tensor with viscous corrections taken in the Israel--Stewart form \cite{Israel:1979wp}. The same decomposition holds for the baryon current,
\begin{equation}\label{eq:currentSplit1}
    J^{\mu}_{\text{total}}(x) = {\cal P}(\tau)\,J^{\mu}_{\text{urqmd}} + \left(1-{\cal P}(\tau)\right)\,J^{\mu}_{\text{hydro}}\,.
\end{equation}

We assume that the rate of thermalization, encoded in the weight function ${\cal P}(\tau)$, depends only on the proper time $\tau$ and not on the transverse coordinates or space-time rapidity; that is, the same weight ${\cal P}(\tau)$ is applied uniformly to every spatial cell on a given $\tau=\text{const}$ hypersurface. The weight function ${\cal P}(\tau)$ satisfies
\begin{equation}
    {\cal P}(\tau_0)=1\,,\quad 0\le {\cal P}(\tau)\le 1\,,
\end{equation}
\begin{equation}\label{eq:Pansatz_boundary2}
    {\cal P}(\tau_{\text{th}})=0\,,\qquad \partial_\tau{\cal P}(\tau_{\text{th}})=0\,.
\end{equation}
In this paper we use the particular ansatz
\begin{equation}\label{eq:Pansatz}
    {\cal P}(\tau) = \left(\frac{\tau_{\text{th}}-\tau}{\tau_{\text{th}}-\tau_0}\right)^{\frac{\tau_{\text{th}}-\tau_0}{\tau_{\text{rel}}}}\,,
\end{equation}
where $\tau_{\text{rel}} < \tau_{\text{th}} - \tau_0$ is another free parameter that sets the rate of the relaxation process.

Since UrQMD conserves energy, momentum, and baryon charge,
\begin{equation}\label{eq:urqmd_conserv}
    \partial_{\mu} T^{\mu\nu}_{\text{urqmd}} = 0\,, \qquad \partial_{\mu}J^{\mu}_{\text{urqmd}} = 0\,,
\end{equation}
and we likewise require the total tensor and current to satisfy the corresponding conservation laws,
\begin{equation}
    \partial_{\mu} T^{\mu\nu}_{\text{total}} = 0\,, \qquad \partial_{\mu}J^{\mu}_{\text{total}} = 0\,,
\end{equation}
substituting the decomposition~(\ref{eq:tensorSplit1})--(\ref{eq:currentSplit1}) into these conditions and using Eq.~(\ref{eq:urqmd_conserv}) yields evolution equations for the hydrodynamic component alone,
\begin{align}\label{eq:hydro_source}
    \partial_\mu \widetilde{T}_{\rm hydro}^{\mu\nu}(x)
    &=
    -T_{\rm urqmd}^{\mu\nu}(x)\,\partial_\mu {\cal P}(\tau),
    \\
    \partial_\mu \widetilde{J}_{\rm hydro}^{\mu}(x)
    &=
    -J_{\rm urqmd}^{\mu}(x)\,\partial_\mu {\cal P}(\tau),
\end{align}
where
\begin{align}
    \widetilde{T}_{\rm hydro}^{\mu\nu}
    &=
    \left[1-{\cal P}(\tau)\right]T_{\rm hydro}^{\mu\nu},
    \\
    \widetilde{J}_{\rm hydro}^{\mu}
    &=
    \left[1-{\cal P}(\tau)\right]J_{\rm hydro}^{\mu}.
\end{align}
The terms on the right-hand side, proportional to $\partial_\mu {\cal P}(\tau)$, thus act as explicit source terms in the hydrodynamic conservation equations, describing the local transfer of energy, momentum, and baryon charge from the nonequilibrium UrQMD component to the equilibrium hydrodynamic component as thermalization proceeds. These source terms are present only during the relaxation stage, and they vanish smoothly after $\tau_{\text{th}}$, and Eqs.~(\ref{eq:hydro_source}) reduce continuously to the standard viscous hydrodynamic conservation equations, $\partial_\mu T_{\rm hydro}^{\mu\nu}=0$ and $\partial_\mu J_{\rm hydro}^{\mu}=0$.





\subsubsection{Hydrodynamic stage}

Once thermalization is achieved, the system evolves according to relativistic viscous hydrodynamics, governed by the conservation of energy-momentum and baryon number.

A key input is the equation of state (EoS), which relates energy density, pressure, and conserved charge densities (baryon number, electric charge, strangeness) to temperature and chemical potentials. The EoS characterizes both the quark-gluon plasma and hadronic phases and incorporates either a smooth crossover or a first-order phase transition. Both cases are considered in this study.

While ideal hydrodynamics assumes vanishing viscosity, realistic modeling includes shear (and sometimes bulk) viscosity. In this work, only shear viscosity is considered, characterized by the ratio $\eta/s$.

When the medium becomes sufficiently dilute, the system transitions to the hadronic cascade via UrQMD, based on a fixed energy density criterion.

\subsubsection{Particlization}

The transition from hydrodynamics to the hadron cascade is performed on a hypersurface of constant energy density $\epsilon_{\text{sw}}$, a free parameter of the model that is fixed independently for each EoS via the spectra calibration of \cite{Rathod:2025gvj} and held fixed across all collision energies studied here. The particlization hypersurface $\Sigma_\mu(x)$ is constructed during the hydrodynamic evolution using the Cornelius routine \cite{huovinen2012, pratt2014}, which, at each hydrodynamic time step, locates the elements of the $\epsilon=\epsilon_{\text{sw}}$ isosurface within each computational cell and outputs them as small hypersurface patches $\Delta\Sigma_\mu$, together with the flow velocity and thermodynamic quantities at each patch. The full particlization hypersurface $\Sigma_\mu(x)$ entering Eq.~(\ref{Cooper-Frye1}) is then assembled by collecting all such patches $\Delta\Sigma_\mu$ over the entire spatial grid and the full duration of the hydrodynamic evolution.

Once the hypersurface $\Sigma_\mu(x)$, is known, we convert the fluid into hadrons by applying the Cooper--Frye prescription \cite{cooper1974} to each hypersurface element $\Delta\Sigma_\mu$,
\begin{equation}
p^0 \frac{d^3 N_i(x)}{d^3 p} = d\Sigma_{\mu}(x)\, p^{\mu}\, f_i\big(x, p\big),
\label{Cooper-Frye1}
\end{equation}
where $i$ labels the hadron species, and Eq.~(\ref{Cooper-Frye1}) is applied to every species included in the hadronic cascade. Since the hypersurface $\Sigma_\mu(x)$ is constructed continuously throughout the hydrodynamic stage, part of it can fall within the relaxation window $\tau_0\le\tau<\tau_{\text{th}}$, where the system is only partially thermalized. For such elements, the distribution function $f_i(x,p)$ inherits the same two-component structure as the total energy-momentum tensor and baryon current, Eq.~(\ref{eq:tensorSplit1}), reflecting the admixture of the nonequilibrium UrQMD component and the equilibrium hydrodynamic one. For hypersurface elements with $\tau\ge\tau_{\text{th}}$, where relaxation is complete, $f_i(x,p)$ reduces to the standard Bose or Fermi equilibrium distribution with viscous corrections. The explicit form of $f_i(x,p)$ in both regimes is given in \cite{Adzhymambetov:2024zzz}, to which we refer the reader for further details.



\subsubsection{Hadron gas cascade}

The hadrons produced at particlization are fed into UrQMD for further evolution. Since UrQMD operates on a constant-time hypersurface $t=$~const, while our particlization surface is not, particles are propagated backward in time (without interactions) to align with UrQMD input requirements. The subsequent evolution, including rescattering and decay, proceeds until $t=400$~fm/$c$, where the final decay of all remaining unstable particles is performed. This time is therefore a technical cutoff of the UrQMD cascade rather than a physical kinetic freeze-out time. The coordinates used to compute the interferometry radii are the emission points of final pions, i.e. their points of last interaction or decay, together with their final momenta.

\subsection{Parameters of the model}

To achieve our goals of describing the HBT-radii at different collision energies, we consider different values of collision energy $\sqrt{s_{NN}}$: 7.7, 11.5, 14.5, 19.6, 27, and 39 GeV. For all these energies, we also consider two different equations of state:  
\begin{enumerate} 
\item The equation of state that implies a cross-over transition from QGP to hadron gas (CO, Chiral EoS) \cite{Steinheimer:2010ib}
\item The equation of state that implies a first-order phase transition (PT, phase-transition EoS) \cite{Kolb:2003dz}
\end{enumerate}
Thus, by comparing the results obtained using different EoS for the same collision energy, we can conclude which transition type leads to a better description of observables: crossover or first-order phase transition. In our model, we have the following free parameters mentioned in the previous section:

\begin{itemize}
\item $R$ $\left[\texttt{fm}\right]$  -- Gaussian smearing parameter used for making UrQMD initial conditions smoother for the relaxation stage.
\item $\tau_{0}$ $\left[\texttt{fm}/c\right]$ -- the starting time of the relaxation stage
\item $\tau_{\text{rel}}$ $\left[\texttt{fm}/c\right]$ -- relaxation time, which defines the speed of relaxation process
\item $\tau_{\text{th}}$ $\left[\texttt{fm}/c\right]$ -- the finishing time of the relaxation stage and, subsequently, the start of the pure hydrodynamic stage
\item $\eta/s$ -- shear viscosity to entropy density ratio
\item $\epsilon_{\text{sw}}$ $\left[\texttt{GeV/fm}^3\right]$ -- energy density corresponding to the transition from the hydrodynamic stage to the hadron gas stage
\end{itemize}

For the sake of simplicity, we take the relaxation time close to its maximal allowed value:
\begin{equation}\label{eq:tauth}
\tau_{\text{rel}} = \tau_{\text{th}} - \tau_0 - \Delta\tau,
\end{equation}
where $\Delta\tau$ is the evolution time step. This choice ensures that the endpoint conditions in Eq. \ref{eq:Pansatz_boundary2} are satisfied. 

In this paper, we use our previous results for the model parameter values \cite{Rathod:2025gvj}. 

\begin{table}[h]
\caption{\label{tab:params}
Parameters of iHKM used for the calculation of interferometry radii.}
\begin{ruledtabular}
\begin{tabular}{ccccccc}
$\sqrt{s_{NN}}$ & EoS & R & $\tau_{0}$ & $\tau_{th}$ & $\eta$ / s & $\epsilon_{\text{sw}}$ \\ \hline
7.7 & PT & 0.5 & 2.7 & 3.3 & 0.08 & 0.35 \\
7.7 & CO & 0.5 & 2.5 & 3.3 & 0.08 & 0.50 \\ 
11.5 & PT & 0.5 & 2.0 & 2.6 & 0.08 & 0.35 \\
11.5 & CO & 0.5 & 1.8 & 2.6 & 0.08 & 0.50 \\ 
14.5 & PT & 0.5 & 1.4 & 2.0 & 0.08 & 0.35 \\
14.5 & CO & 0.5 & 1.3 & 2.3 & 0.08 & 0.50 \\
19.6 & PT & 0.5 & 1.0 & 1.6 & 0.08 & 0.35 \\ 
19.6 & CO & 0.5 & 1.0 & 1.6 & 0.08 & 0.50 \\
27.0 & PT & 0.5 & 0.8 & 1.4 & 0.08 & 0.35 \\ 
27.0 & CO & 0.5 & 0.8 & 1.4 & 0.08 & 0.50 \\
39.0 & PT & 0.5 & 0.6 & 1.2 & 0.08 & 0.35 \\
39.0 & CO & 0.5 & 0.6 & 1.6 & 0.08 & 0.50 \\ 
\end{tabular}
\end{ruledtabular}
\end{table}

\subsection{Interferometry radii}

The method of stellar intensity interferometry of Hanbury-Brown and Twiss was proposed in the late 1950s by astronomers Hanbury-Brown and Twiss for measuring the angular size of stars.\cite{HBT} Later, in the 1970s, it was successfully transformed into the two-particle correlation femtoscopy/interferometry analysis of the homogeneity lengths in the matter formed in nuclear (or particle) collisions \cite{Kopylov1974, Makhlin:1988anm}.  The similarity and principal difference between HBT and correlation femtoscopy methods one can read in detail in \cite{Sinyukov:2013Femtoscopy}. The correlation femtoscopy/interferometry is a useful tool for studying the space-time structure of the particle-emitting source, see, for example, \cite{Heinz:1996nuclth, heinz99, Wiedemann:1999PR}. 

Interferometry radii should be considered more as homogeneity regions than the real geometric sizes of the emitting source.\cite{Akkelin:1996sg, Sinyukov:1996_NPA610} Mathematically, they are obtained from the two-particle momentum correlation function:
\begin{equation}
C(\vec{p}_1, \vec{p}_2) = \frac{E_1E_2dN/(d^3\vec{p}_1d^3\vec{p}_2)}{(E_1dN/(d^3\vec{p}_1))(E_2dN/(d^3\vec{p}_2))}
\end{equation}

After some transformations, it can be written in terms of the emission function $S(x,p)$ \cite{heinz99}
\begin{equation}
C(\vec{q}, \vec{p}) = 1\pm \frac{\mid \int d^4x S(x,p)e^{iqx}\mid ^ 2}{\int d^4x S(x,p+\frac{1}{2}q) \int d^4y S(y,p-\frac{1}{2}q)}. 
\end{equation}


where pair relative momenta $q = p_1 - p_2$, pair average momenta $p = \frac{p_1 + p_2}{2}$. Then, applying mass shell and smoothness approximation \cite{heinz99}, this transforms to

\begin{equation}
C(\vec{q}, \vec{p}) \approx 1\pm\left|\frac{\int d^4x S(x,p)e^{iqx}}{\int d^4x S(x,p)}\right| ^ 2 = 1\pm \left| \langle e^{iqx}(p) \rangle \right| ^ 2
\end{equation}

In the UrQMD event generator, this equation has the form 
\begin{equation}
\scalebox{1.2}{$C(\vec{q}) = \frac{\sum\limits_{i\neq j}\delta_\Delta(\vec{q}-\vec{p}_i+\vec{p}_j)(1+\cos(\vec{p}_j-\vec{p}_i)(\vec{x}_j-\vec{x}_i))}{\sum\limits_{i\neq j}\delta_\Delta(\vec{q}-\vec{p}_i+\vec{p}_j)},$}
\end{equation}
$\delta_\Delta(\vec{x})=1$ if $\mid \vec{x} \mid < \Delta/2$ and 0 otherwise, where $\Delta $ is the size of the bin in histogram. The correlation function histograms in this form for different pair relative momenta $\vec{q}$ in the LCMS system are fitted with Gaussians for each $\vec{p}$:

\begin{multline}\label{eq:corr_fit}
C(\vec{q}) = 1 + \lambda  \exp \left( -R^2_{out}q^2_{out} - \right.\\ \left. R^2_{side}q^2_{side} - R^2_{long}q^2_{long} \right)
\end{multline}
The interferometry radii $R_{out}(p_T)$, $R_{side}(p_T)$, $R_{long}(p_T)$ and the suppression parameter $\lambda$ are extracted from this fit. In this work, we calculate the dependence of interferometry radii on transverse mass $m_T = \sqrt{m^2+p_T^2}$ to compare it with experimental data \cite{STAR:2015}.

\begin{figure*}
\includegraphics[width=0.49\textwidth]{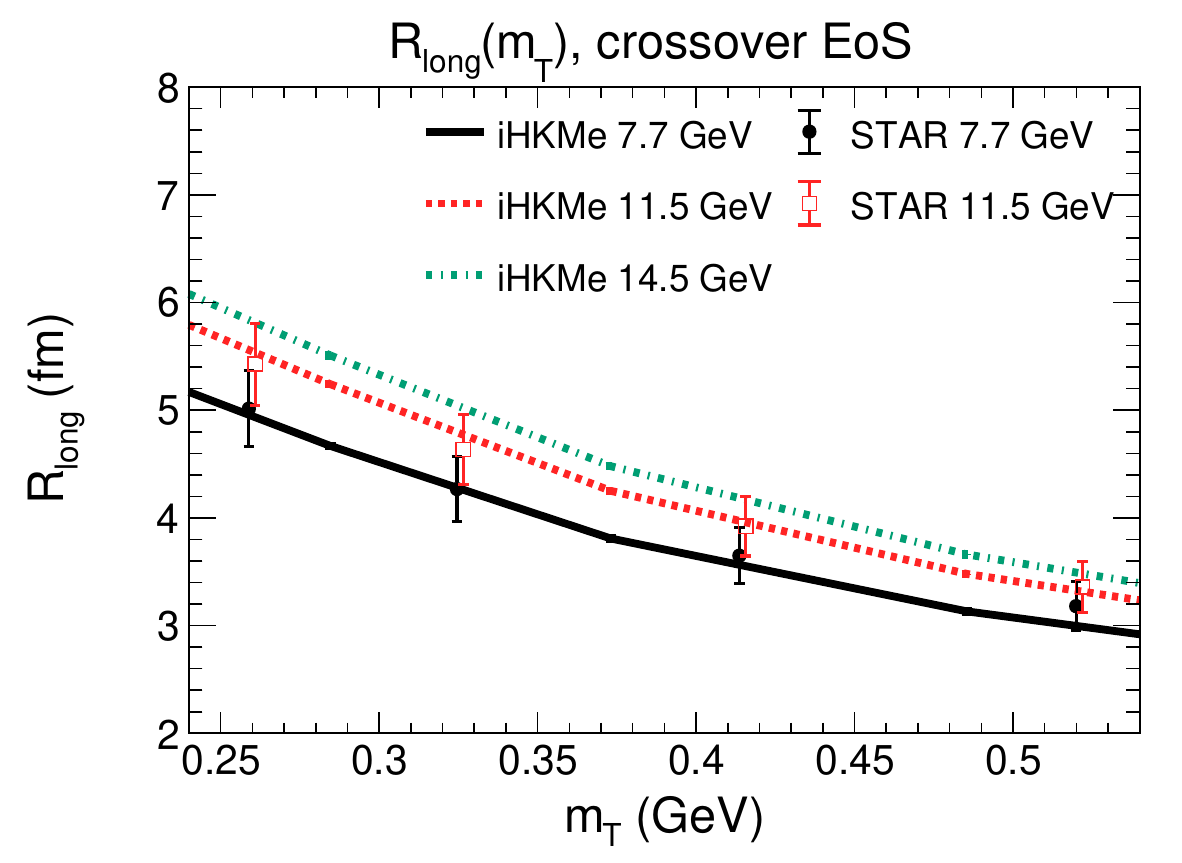}
\includegraphics[width=0.49\textwidth]{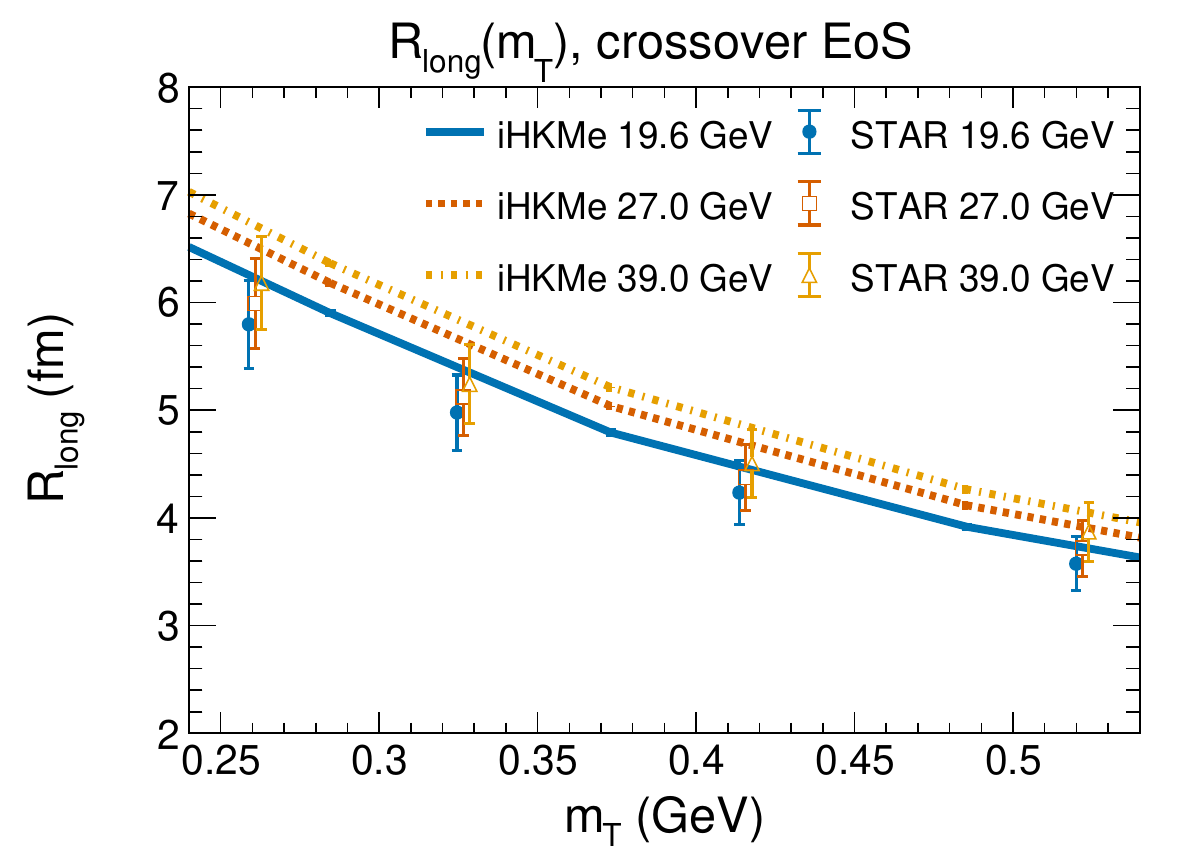}
\includegraphics[width=0.49\textwidth]{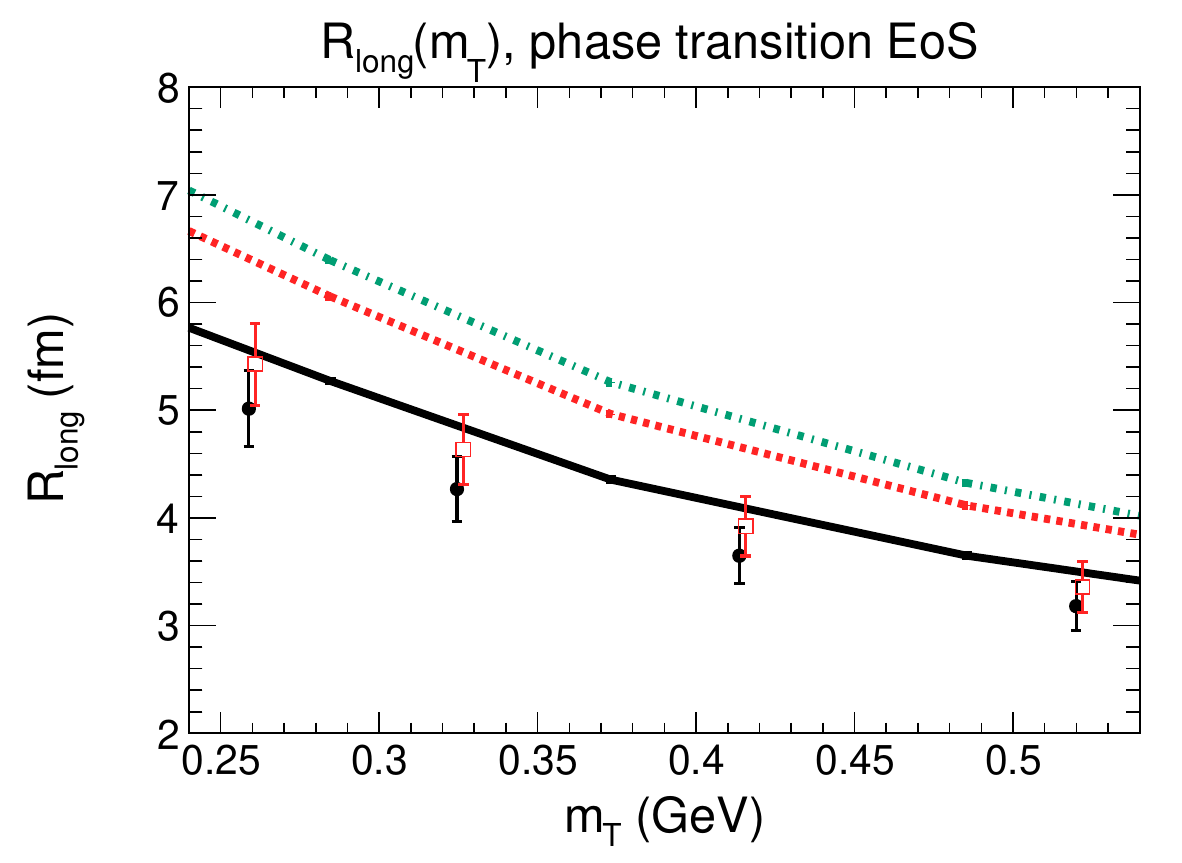}
\includegraphics[width=0.49\textwidth]{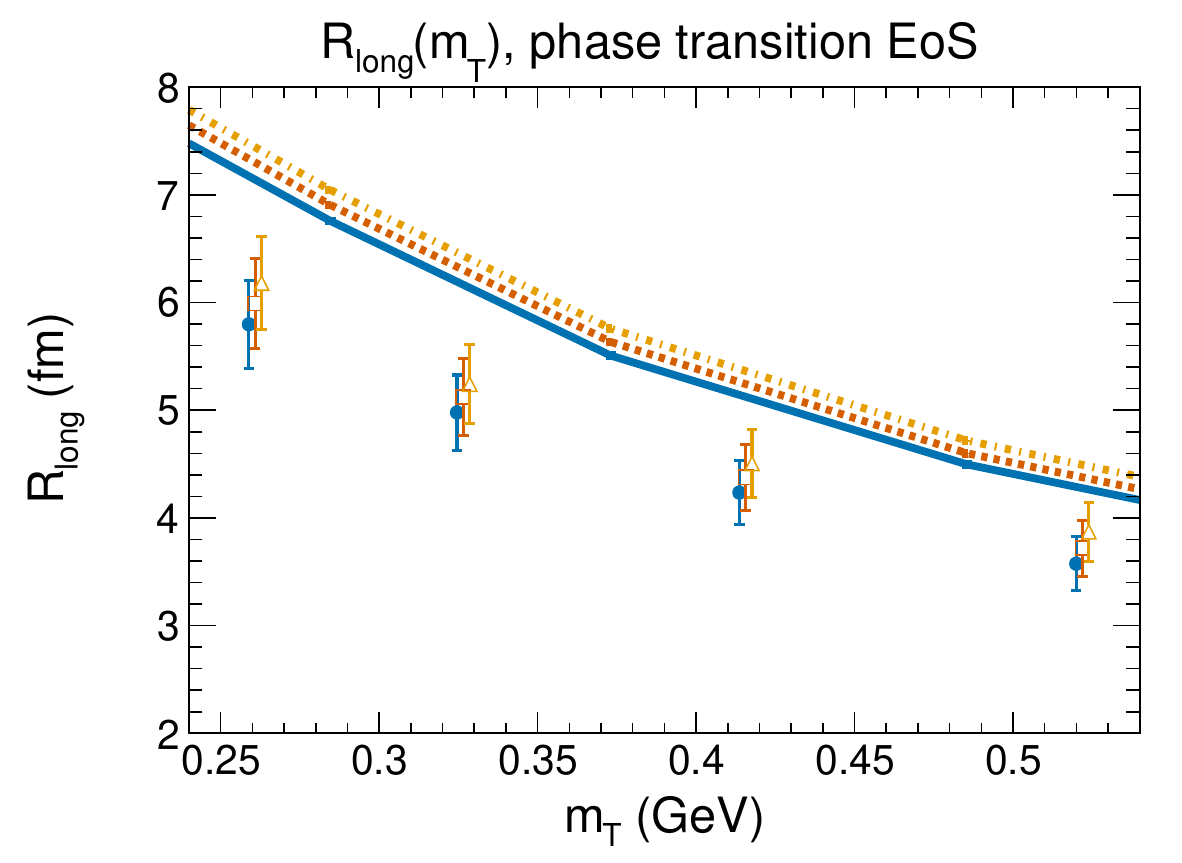}
\caption{Calculated $R_{\rm long}$ radii for $\pi^{-}\pi^{-}$ pairs in the 5\% most central Au+Au collisions over the energy range $\sqrt{s_{NN}}=7.7$--$39.0$~GeV. Results obtained with the chiral crossover EoS and the first-order phase-transition EoS are shown for comparison. The upper and lower rows correspond to the crossover and first-order-transition EoSs, respectively; the left and right columns show the lower- and higher-energy groups. The available experimental data are taken from Ref.~\cite{STAR:2015}. The line and marker conventions shown in the upper panels also apply to the corresponding lower panels.}
\label{fig:R_long}
\end{figure*}

%% file: sections/3_results.tex
For each parameter set, we perform 500 hydrodynamic simulations. From each simulation, we sample 200 UrQMD afterburner events, resulting in a total of 100,000 events per parameter set. All calculations in this study are carried out for the $5\%$ most central collisions. Centrality is defined based on the multiplicity of charged particles. However, due to the limited number of hydrodynamic simulations, it is not feasible to define centrality classes with sufficient statistical precision directly from hydrodynamic data. Therefore, for each collision energy, we first generate 20,000 standalone UrQMD events to define centrality classes. We then select the top $5\%$ most central events as initial conditions for the iHKM pre-equilibrium stage and the subsequent hydrodynamic evolution. In order to have enough statistics for pion pairs of close momentum, a batch of events is taken to calculate the two-particle correlation function.

\begin{figure*}
\includegraphics[width=0.49\textwidth]{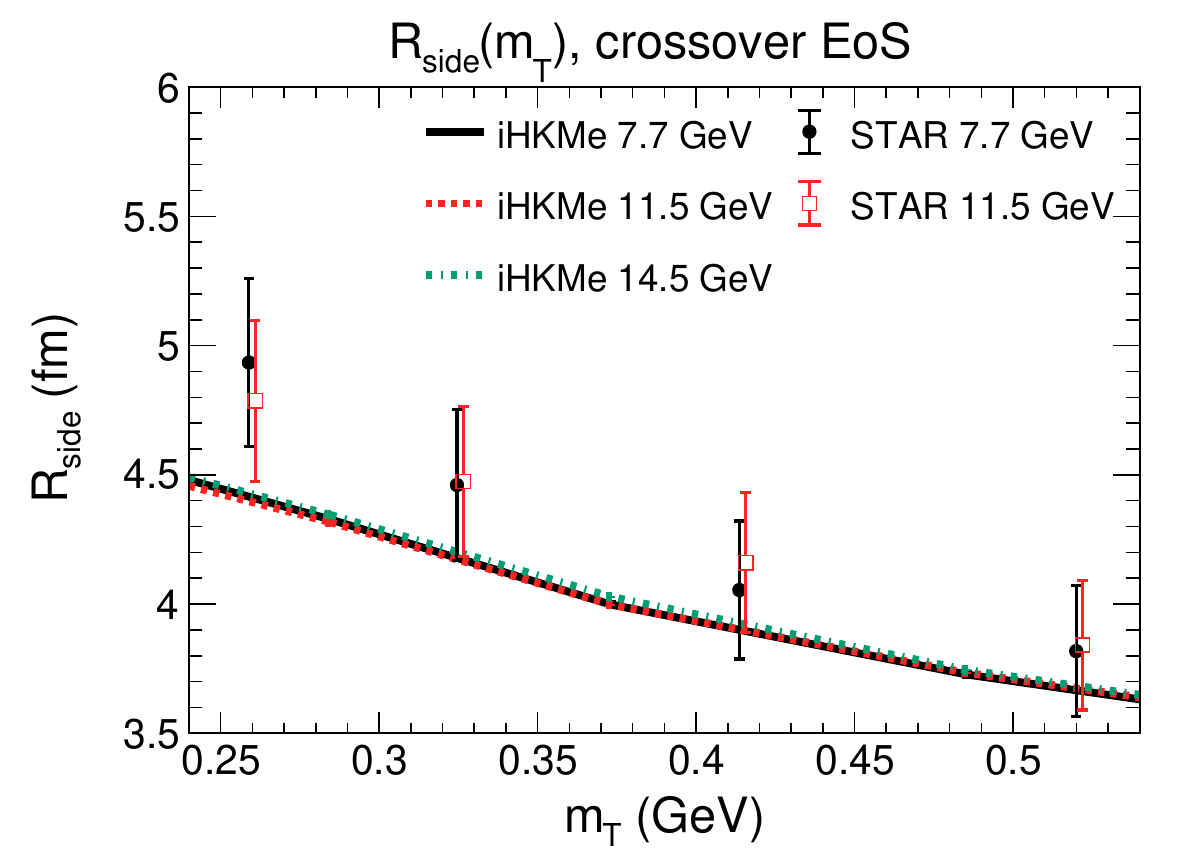}
\includegraphics[width=0.49\textwidth]{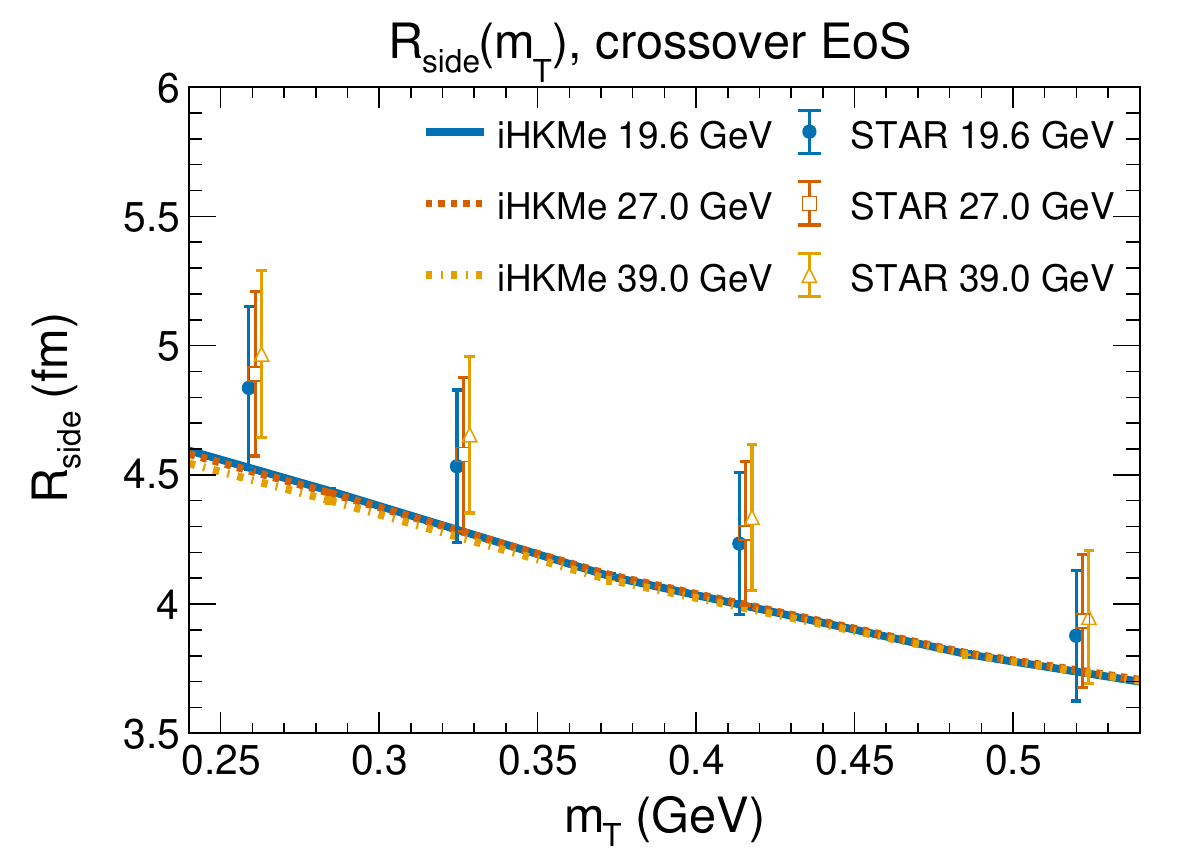}
\includegraphics[width=0.49\textwidth]{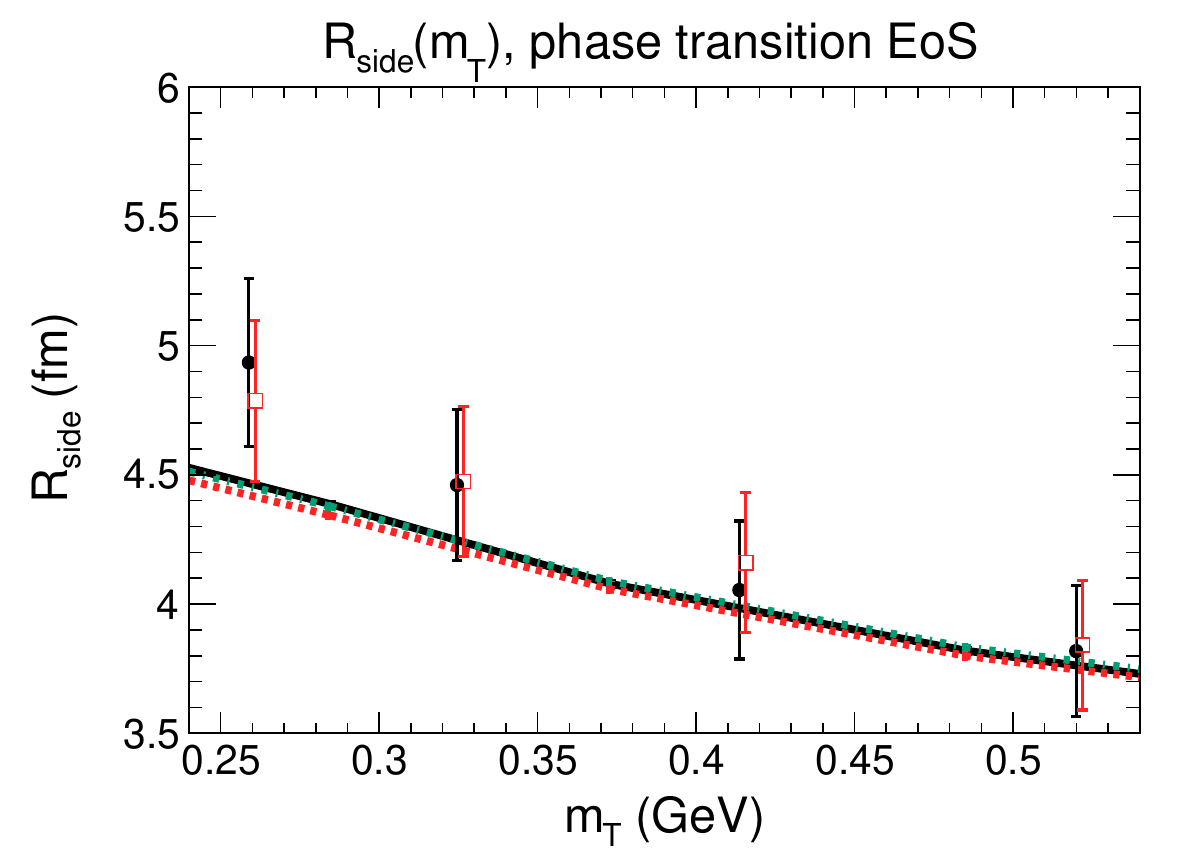}
\includegraphics[width=0.49\textwidth]{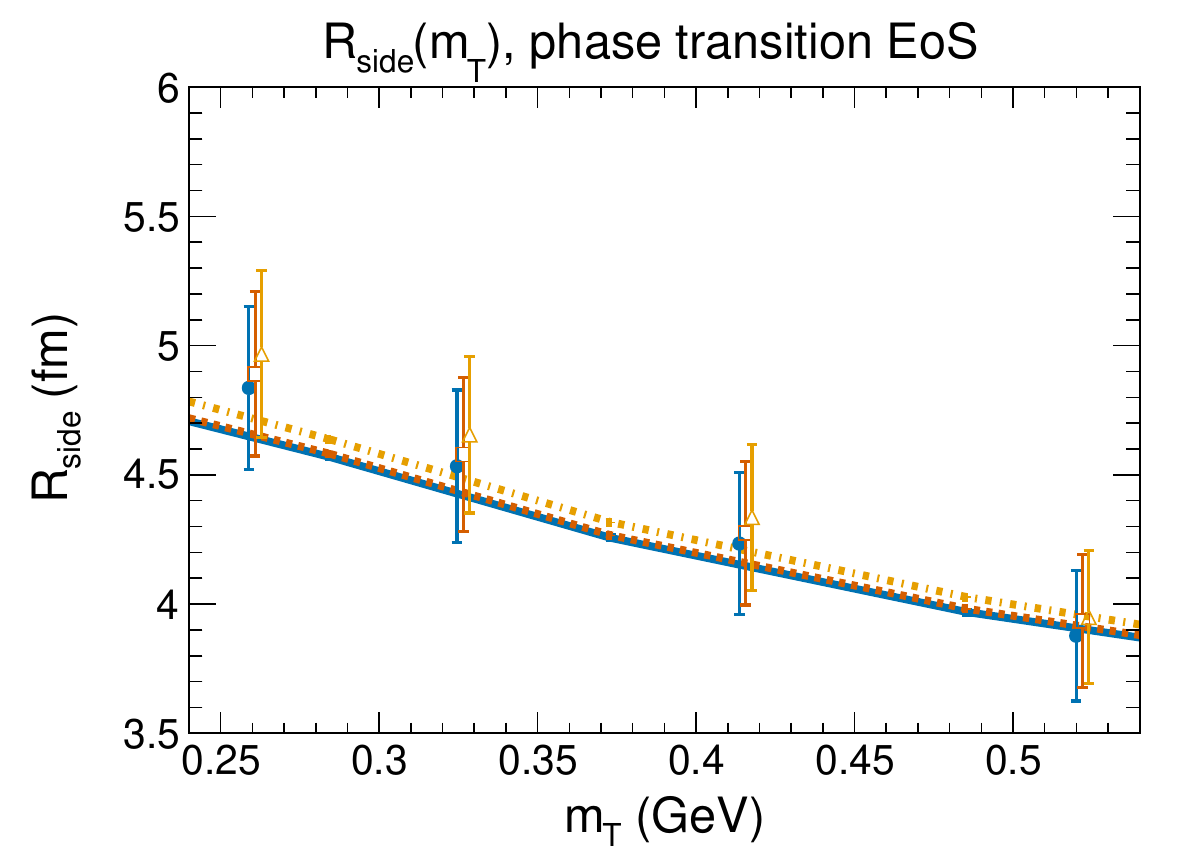}
\caption{Calculated $R_{\rm side}$ radii for $\pi^{-}\pi^{-}$ pairs in the 5\% most central Au+Au collisions over the energy range $\sqrt{s_{NN}}=7.7$--$39.0$~GeV. Results obtained with the chiral crossover EoS and the first-order phase-transition EoS are shown for comparison. The upper and lower rows correspond to the crossover and first-order-transition EoSs, respectively; the left and right columns show the lower- and higher-energy groups. The available experimental data are taken from Ref.~\cite{STAR:2015}. The line and marker conventions shown in the upper panels also apply to the corresponding lower panels.}
\label{fig:R_side}
\end{figure*}

The resulting HBT radii for lower collision energies ($\sqrt{s_{NN}} = 7.7-14.5$~GeV) using Chiral (crossover) and first-order phase transition EoS are presented on Figs. \ref{fig:R_long} - \ref{fig:R_side} (left plots). The experimental error bars include the statistical and systematic uncertainties combined in quadrature. The statistical uncertainties of the calculated radii are smaller than the line width and are therefore not shown. Although experimental data at $\sqrt{s_{NN}}=14.5$~GeV are not available for comparison, the corresponding model results are included for completeness and to examine the predicted energy dependence of the HBT radii. From these figures, we observe that the Chiral EoS provides a slightly better overall description of the HBT radii than the first-order phase transition EoS. The performance of the first-order EoS for $R_{\rm long}$ improves as the collision energy decreases, while no such improvement is seen for $R_{\rm out}$.

As in the STAR data, our results for the \textit{long} HBT radii show an increasing trend with collision energy. This behavior is commonly attributed to an extended system lifetime at higher energies, driven by an increase in initial energy density \cite{Akkelin:1996sg}. In contrast, the experimental \textit{side} and \textit{out} HBT radii exhibit minimal dependence on collision energy. Although our model predicts slightly varying—but closely matching—values for these radii, it reproduces their overall energy independence.

At higher collision energies, we also consider both the chiral (crossover) and the first-order phase transition equations of state. The corresponding results are presented on the right parts of the same Figs. \ref{fig:R_long} - \ref{fig:R_side}. In this energy regime, a clear difference between the two scenarios is observed: while the chiral EoS provides a reasonable description of the experimental data, the first-order phase transition EoS leads to a noticeable deviation, particularly in the \textit{long} interferometry radius. Specifically, the phase transition scenario predicts a significantly stronger increase of $R_{\textit{long}}$ with collision energy than is supported by the data. This behavior indicates that the model with a first-order phase transition becomes less consistent with experimental observations at higher energies and supports the expectation that the transition in this regime is a smooth crossover. The chiral EoS results in this region are therefore more consistent with the experimental trends, although a slight overestimation of the $R_{\textit{long}}$ growth with energy remains.

\subsection{Sensitivity to model parameters}

To examine the sensitivity of the calculated interferometry radii to the
choice of model parameters, we performed additional calculations for the
5\% most central Au+Au collisions at
$\sqrt{s_{NN}}=11.5$~GeV using the crossover EoS. Starting from the
baseline parameter set given in Table~\ref{tab:params}, one independent
parameter was varied at a time, while the remaining independent parameters
were kept unchanged. Since $\tau_{\mathrm{th}}$ is related to
$\tau_0$ and $\tau_{\mathrm{rel}}$ through
Eq.~\eqref{eq:tauth}, variations of either of these parameters also result
in a corresponding change of $\tau_{\mathrm{th}}$.

\begin{figure*}
\includegraphics[width=0.49\textwidth]{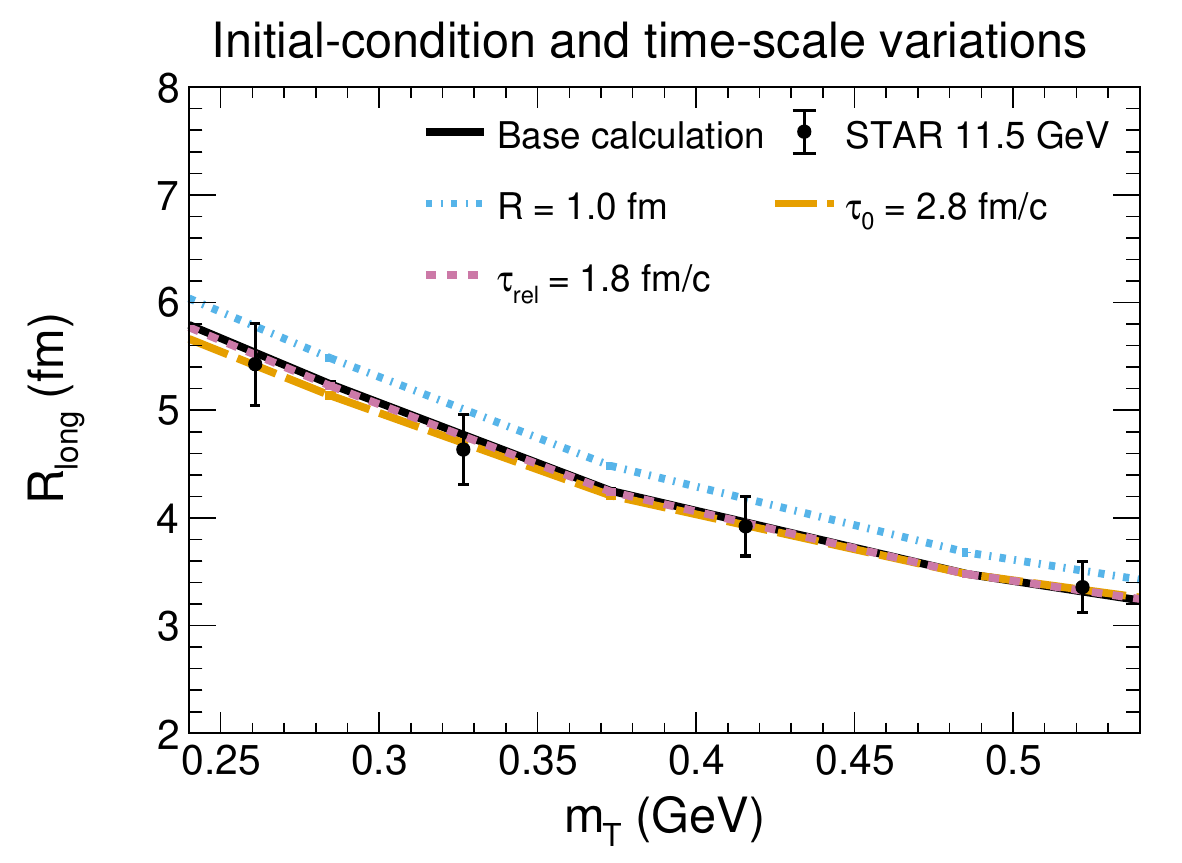}
\includegraphics[width=0.49\textwidth]{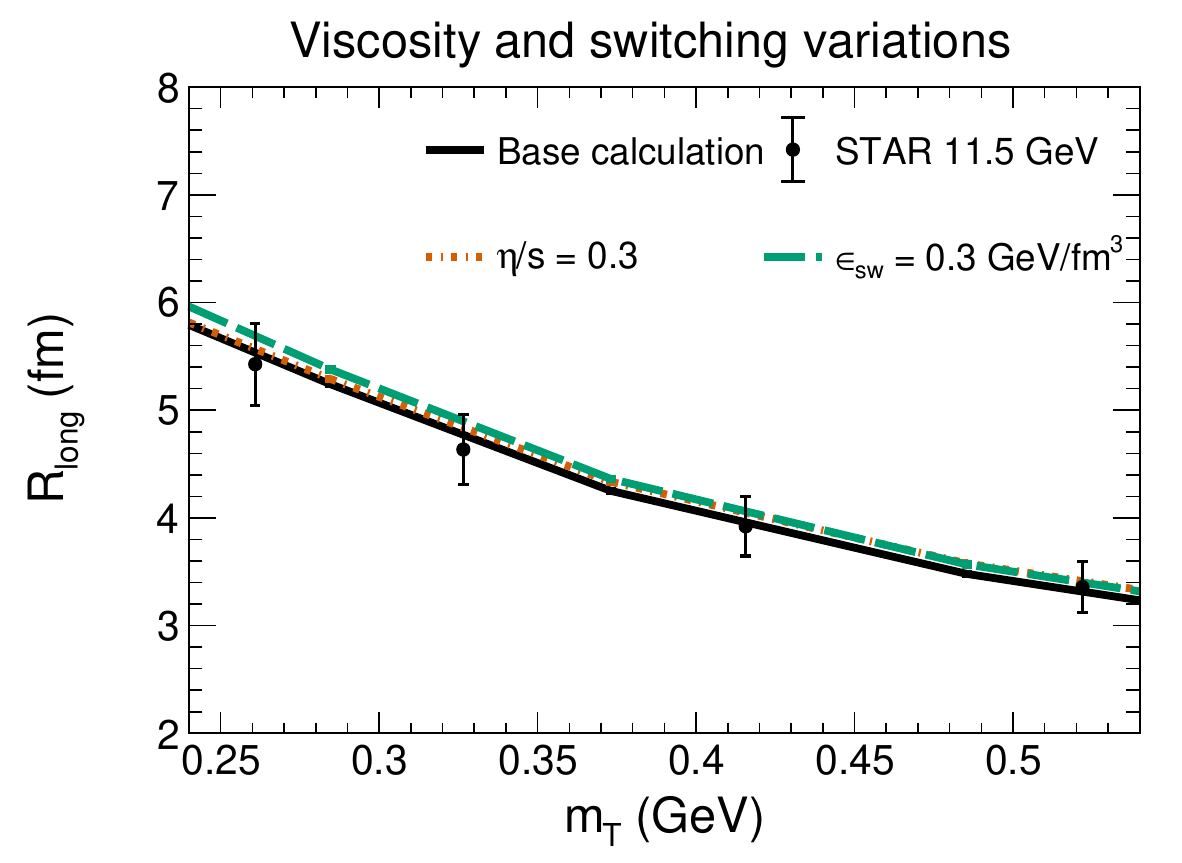}
\includegraphics[width=0.49\textwidth]{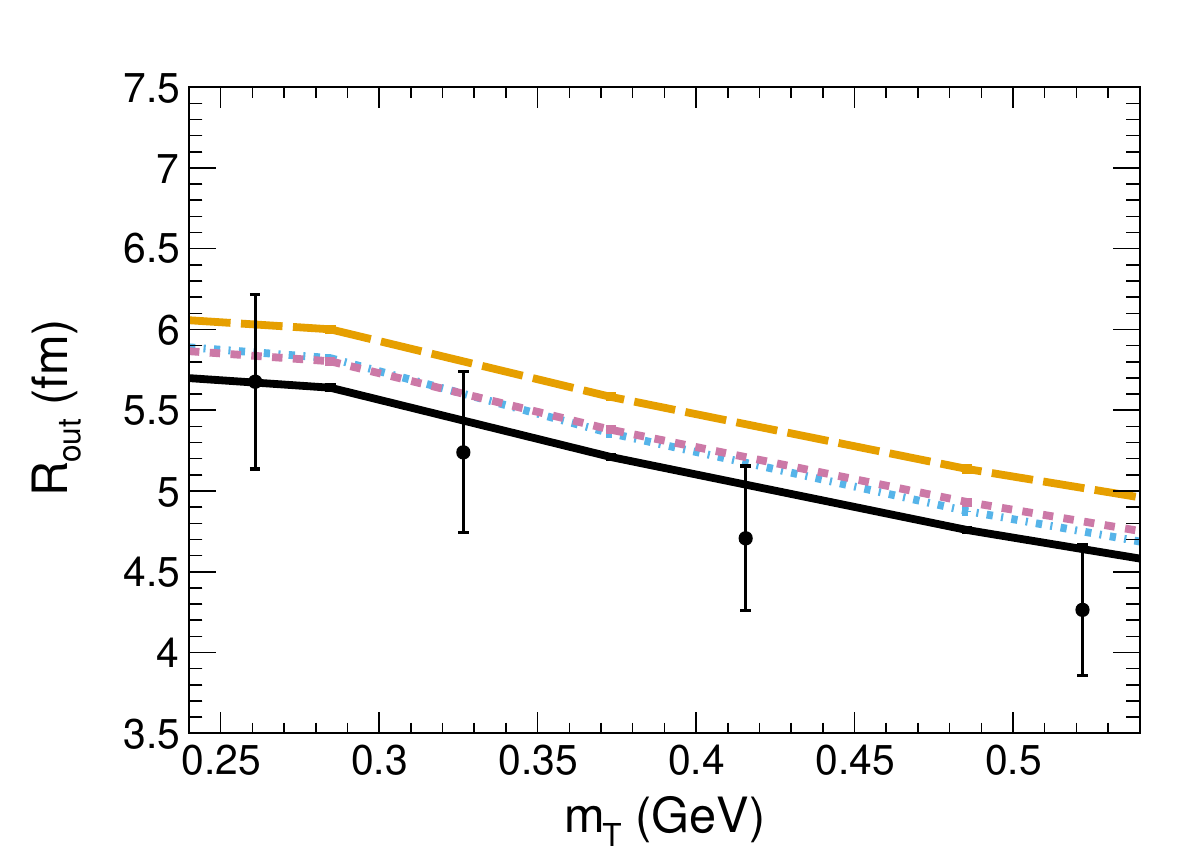}
\includegraphics[width=0.49\textwidth]{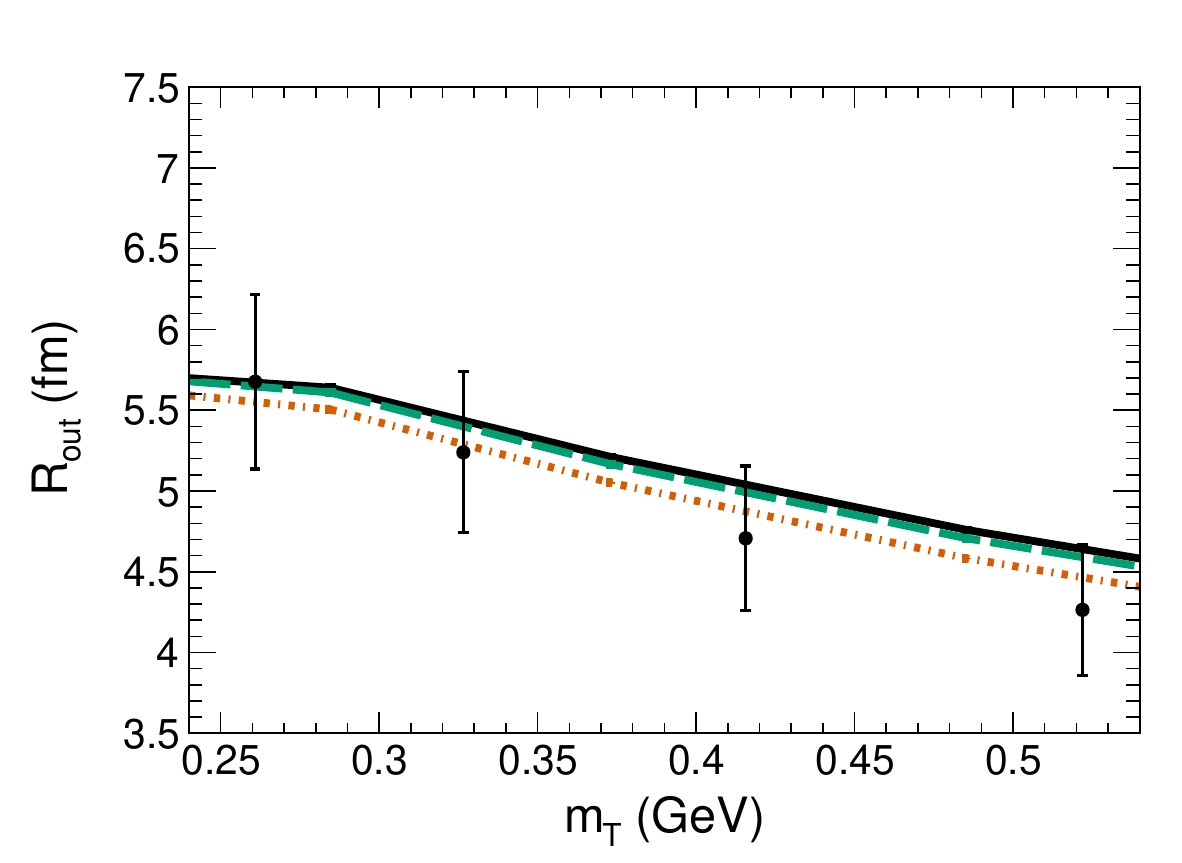}
\includegraphics[width=0.49\textwidth]{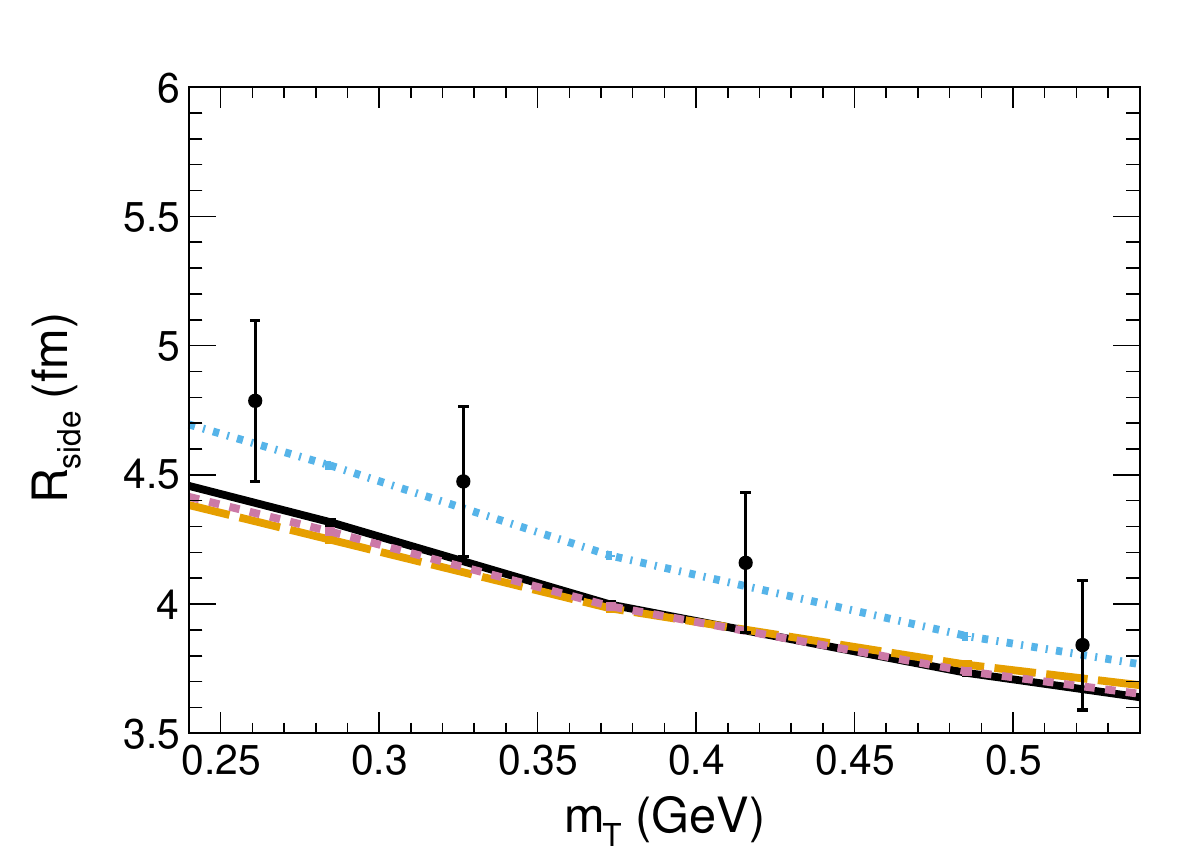}
\includegraphics[width=0.49\textwidth]{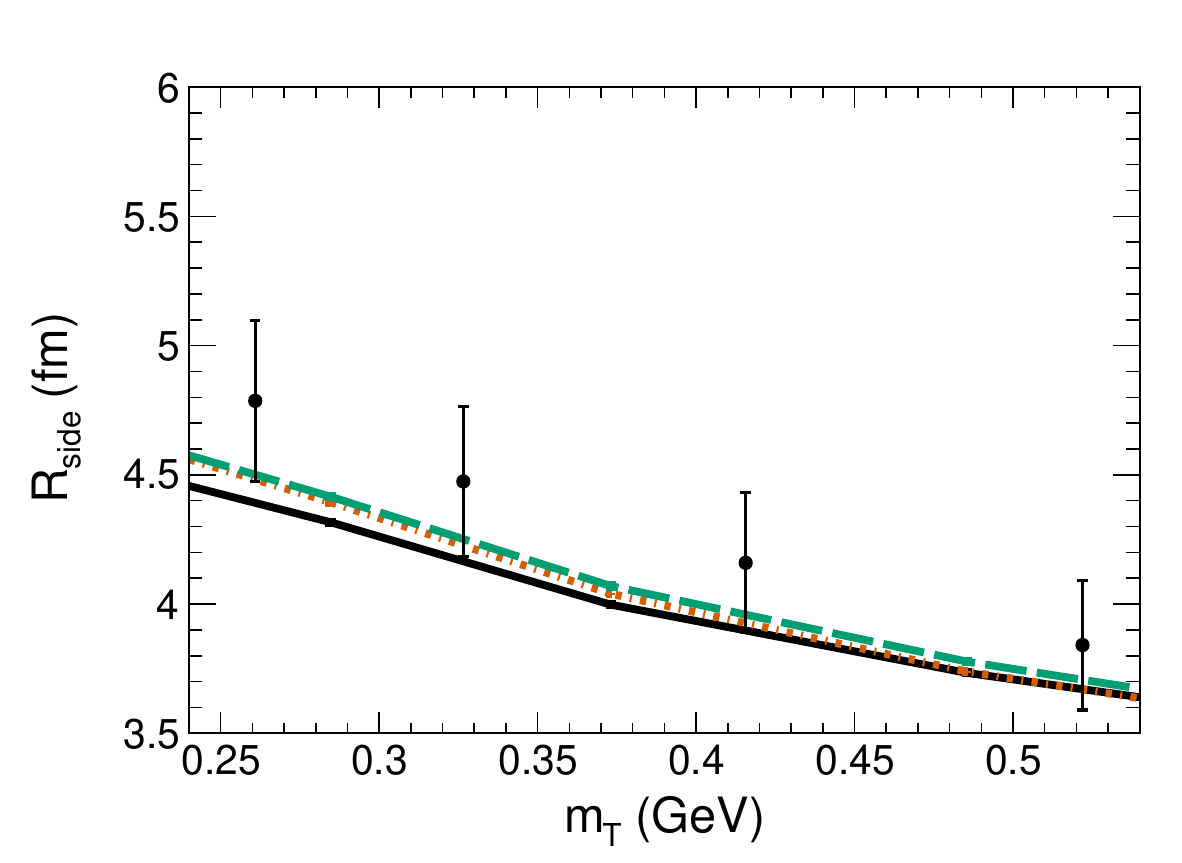}
\caption{Sensitivity of the calculated HBT radii $R_{\rm long}$, $R_{\rm out}$, and $R_{\rm side}$ for $\pi^{-}\pi^{-}$ pairs in the 5\% most central Au+Au collisions at $\sqrt{s_{NN}}=11.5$~GeV, obtained with the chiral crossover EoS. The solid black curves represent the base calculation. The left column shows the effects of varying the initial-condition Gaussian smearing radius $R$, the relaxation-time parameter $\tau_{\rm rel}$, and the initial time $\tau_{0}$, while the right column shows the effects of varying the specific shear viscosity $\eta/s$ and the switching energy density $e_{\rm sw}$. Each independent parameter is varied separately while the other independent parameters are held fixed; when $\tau_{0}$ or $\tau_{\rm rel}$ is varied, $\tau_{\rm th}$ changes consistently with Eq.~\ref{eq:tauth}. The experimental data are taken from Ref.~\cite{STAR:2015}. The line and marker conventions shown in the upper panels also apply to the corresponding lower panels.}
\label{fig:sensitivity_tests}
\end{figure*}

The initial state and pre-equilibrium variations, shown in the left panels
of Fig.~\ref{fig:sensitivity_tests}, include an increase of the Gaussian
smearing radius from $R=0.5$ to $1.0$~fm, an increase of the relaxation
time from $\tau_{\mathrm{rel}}=0.8$ to $1.8$~fm/$c$, and an increase of
the relaxation stage starting time from $\tau_0=1.8$ to
$2.8$~fm/$c$. The sensitivity to the specific shear viscosity and the
particlization criterion is presented in the right panels of
Fig.~\ref{fig:sensitivity_tests}, where $\eta/s$ was increased from
$0.08$ to $0.30$ and the switching energy density was reduced from
$\epsilon_{\mathrm{sw}}=0.50$ to $0.30$~GeV/fm$^{3}$.

The considered variations produce moderate changes in the absolute values
of the radii without substantially modifying their qualitative
$m_T$ dependence. The most visible effects are associated with the
initial state and pre-equilibrium parameters. Increasing $R$ generally
increases $R_{\mathrm{long}}$ and $R_{\mathrm{side}}$, whereas increasing
$\tau_0$ increases $R_{\mathrm{out}}$ and decreases
$R_{\mathrm{long}}$ and $R_{\mathrm{side}}$. By comparison, variations
of $\eta/s$ and $\epsilon_{\mathrm{sw}}$ produce smaller changes.
Overall, this representative sensitivity study indicates that the qualitative behavior of the calculated HBT radii remains stable under the considered one-at-a-time parameter variations. Although some variations move
individual radii slightly closer to the experimental data, the baseline
parameter set was not fitted to the HBT observables, but was adopted from
the independent calibration to particle spectra performed in
Ref.~\cite{Rathod:2025gvj}.

\begin{figure*}
\includegraphics[width=0.49\textwidth]{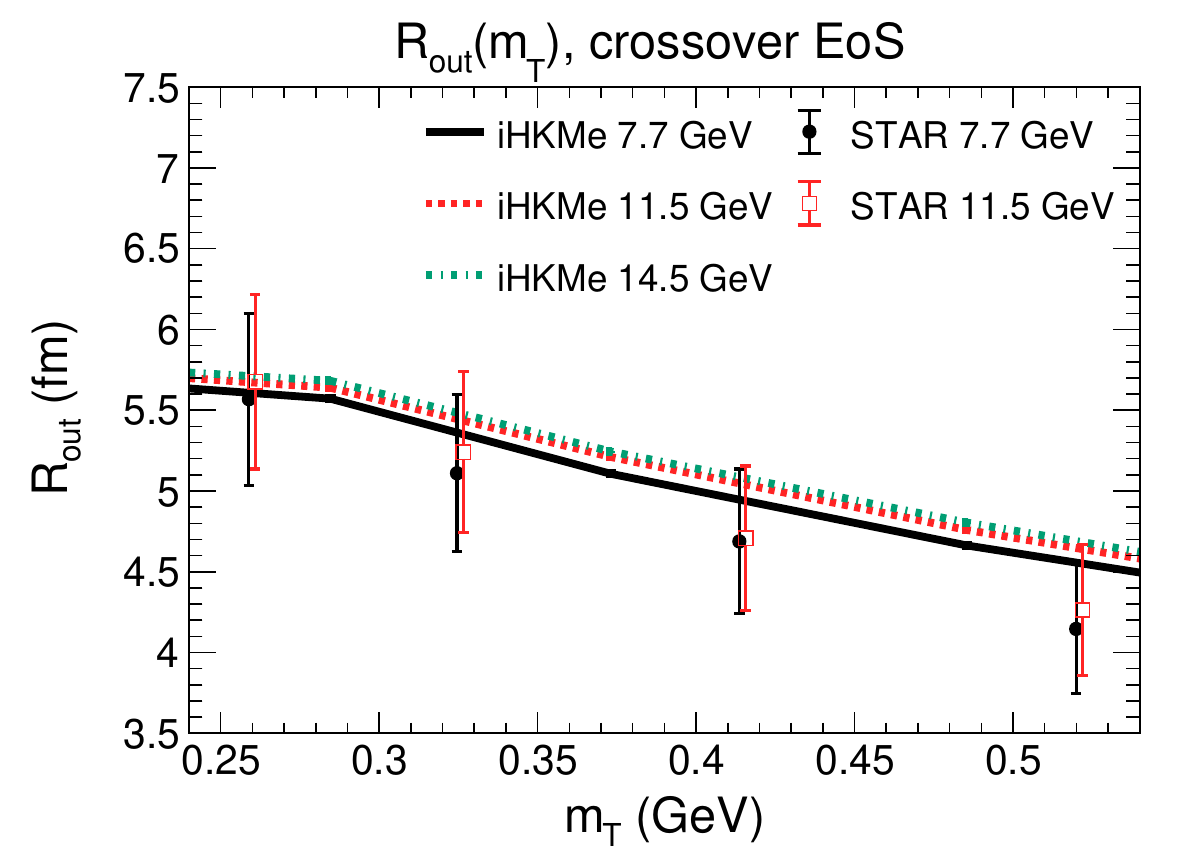}
\includegraphics[width=0.49\textwidth]{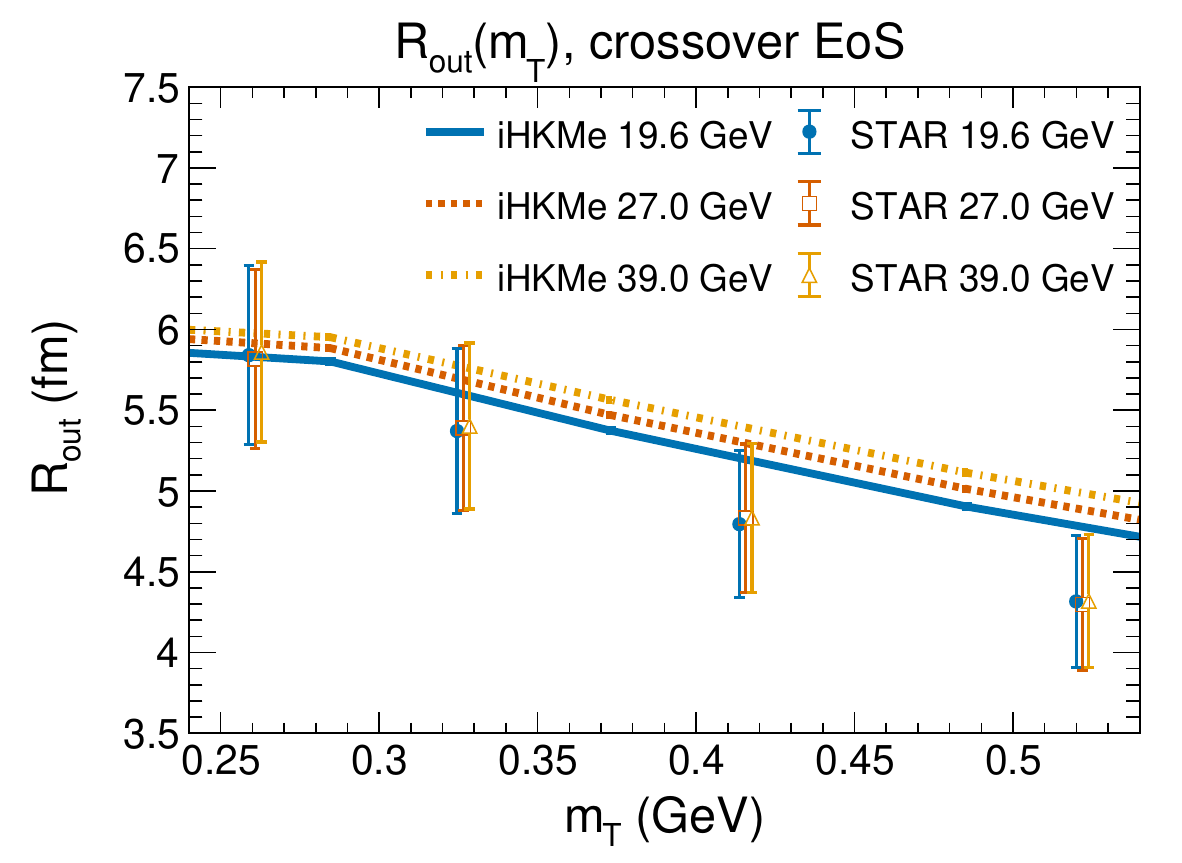}
\includegraphics[width=0.49\textwidth]{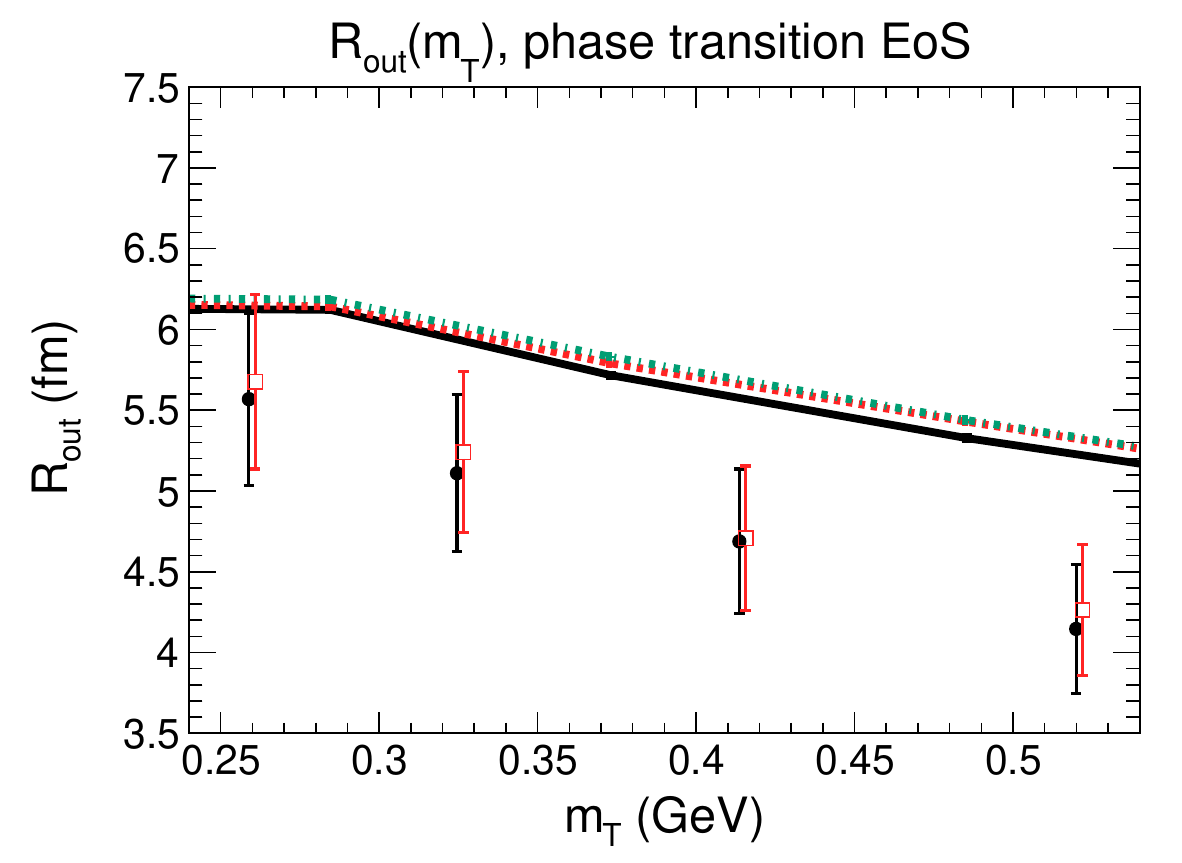}
\includegraphics[width=0.49\textwidth]{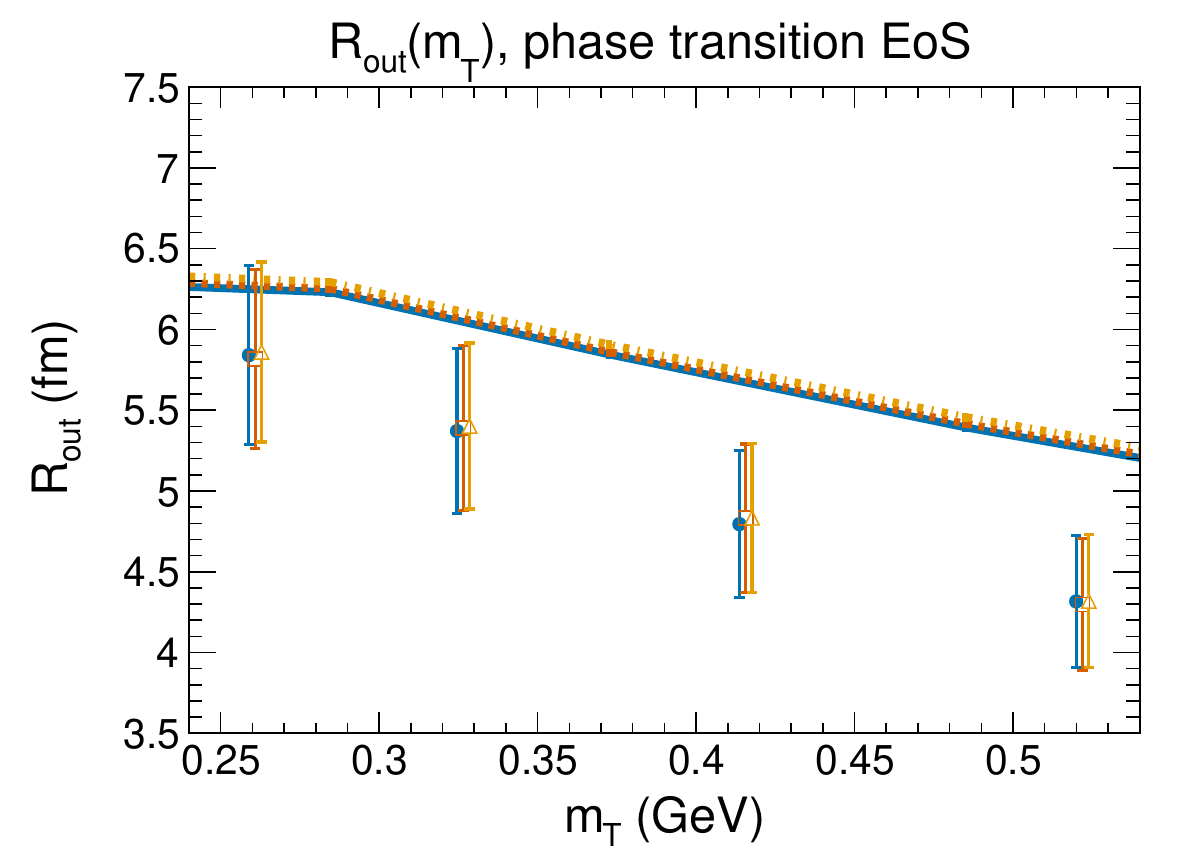}
\caption{Calculated $R_{\rm out}$ radii for $\pi^{-}\pi^{-}$ pairs in the 5\% most central Au+Au collisions over the energy range $\sqrt{s_{NN}}=7.7$--$39.0$~GeV. Results obtained with the chiral crossover EoS and the first-order phase-transition EoS are shown for comparison. The upper and lower rows correspond to the crossover and first-order-transition EoSs, respectively; the left and right columns show the lower- and higher-energy groups. The available experimental data are taken from Ref.~\cite{STAR:2015}. The line and marker conventions shown in the upper panels also apply to the corresponding lower panels.}
\label{fig:R_out}
\end{figure*}

\clearpage

%% file: sections/4_summary.tex
In this work, we applied the recently extended integrated HydroKinetic Model to describe femtoscopic observables in relativistic nucleus–nucleus collisions at the intermediate energies of the RHIC Beam Energy Scan program. The updated model accounts for essential features of lower-energy collisions, including significant baryon stopping, prolonged nuclear overlap times, and the possible partial thermalization of matter. Two scenarios for the equation of state of strongly interacting matter—a smooth crossover and a first-order phase transition—were investigated. The calculated femtoscopic observables, particularly the interferometry radii over the broad energy range from 7.7 to 39 GeV, demonstrate the model’s capability to reproduce key experimental trends and offer insights into the QCD phase structure at intermediate energies.

Comparison with experimental data demonstrates that the extended iHKM framework provides a satisfactory description of the space-time structure of the particle-emitting source across the entire energy range. The results show that the interferometry radii are sensitive to the choice of EoS, offering potential insights into the nature of the QCD phase transition at intermediate collision energies.

We find that the crossover EoS yields a better description of the interferometry radii, especially at higher BES energies, while at lower energies, such as 7.7 GeV, the difference between the two equations of state becomes less pronounced for {\it long} radii. However, in the lower energy range the Chiral EoS still provides a slightly better overall description of the HBT radii. 

The {\it side} and {\it out} radii show minimal dependence on collision energy, whereas the {\it long} radius exhibits a clear increasing trend with energy.

Further refinements of the model and systematic studies at lowest relativistic energies are planned to deepen our understanding of the properties of strongly interacting matter created in these collisions.

\begin{acknowledgments}
H.Z.'s work was supported by the Grant of National Science Centre, Poland, No: 2021/41/B/ST2/02409 and 2020/38/E/ST2/00019. The research of Yu. S. was partially funded by IDUB-POB projects granted by WUT, Warsaw, Poland, and was supported in part by the ExtreMe Matter Institute EMMI at the GSI Helmholtzzentrum für Schwerionenforschung GmbH, Darmstadt, Germany. M.A. and V. N. gratefully acknowledge support from the National Research Foundation of Ukraine under grant 2025.07/0050.
\end{acknowledgments}